\documentclass [12pt]{article}

\begin{document}
\begin{center}
A STRINGY COMPOSITE GRAVITON AND THE COSMOLOGICAL CONSTANT PROBLEM
\end{center}

\begin{center}
David L. Henty
\end{center}

\begin{center}
Department of Physics and Astronomy,
\end{center}

\begin{center}
University of British Columbia,
\end{center}

\begin{center}
Vancouver, British Columbia V6T 1Z1
\end{center}

\begin{center}
Email address: henty@physics.ubc.ca
\end{center}

\begin{center}
ABSTRACT
\end{center}

A heuristic model of a composite graviton is presented motivated by open 
string field theory. The model simply assumes a composite closed string 
sector, world sheet conformal invariance and the observed open string 
states, i.e., standard model particles, as an input. A UV to IR map from the open 
string to the closed string sectors naturally emerges. The IR cutoff of the 
lightest observed fermion loops gets mapped to a UV cutoff for the scalar 
mode of the graviton/dilaton. This provides a UV cutoff for gravity via the 
dilaton coupling to G. Based on recent estimates of the lightest fermion 
(neutrino) mass this gives an energy cutoff for the graviton near the 
current estimates of the dark energy density of 10$^{-3}$ eV. 

\newpage 
1. INTRODUCTION

The cosmological constant problem is arguably the greatest problem currently 
facing string theory or any other approach to quantum gravity. [1] The 
reason this problem is so significant is that involves energy scales which 
are believed to be well understood by current theory. This has been driven 
home by the recent cosmological expansion data giving an effective 
cosmological constant or dark energy density of about 10$^{-3}$ electron 
volts. The cosmological constant problem is therefore not a Planck energy 
scale problem but rather a problem involving energy scales on the order of 
10$^{-3}$ electron volts, an energy scale we thought we understood 
completely. On the other hand, if the cosmological constant problem truly is 
solved at the Planck energy scale then it becomes a fine tuning problem of 
unbelievable precision, i.e., a precision of 10$^{-120}$! The current 
situation is so desperate that to many the anthropic principle seems to be 
the only way out. [2]

Perhaps as a result of this, the cosmological constant problem has recently 
been examined in a number of qualitative and/or phenomenological approaches 
to see if it is at least \textbf{possible} to address the problem outside of 
the anthropic principle. [3]\footnote{ The present paper is presented in 
this spirit as heuristic and admittedly speculative ideas previously limited 
to private communications are being vetted for a wider audience.} For 
example, it has recently been suggested by more than one author that the 
possible solution to the cosmological constant problem is in the nature of a 
composite graviton although the specific model for such a composite 
structure has generally not been discussed.\footnote{ E.g., see [3], A. Zee 
and R. Sundrum.} In unrelated work it has for some time been conjectured 
that a composite graviton could naturally arise in open string field theory. 
[4][6] Basically, the graviton propagator is derived from the open string 
loops. At present the technical details of such an approach are unresolved. 
Nonetheless, it seems quite possible that while this type of construction 
may be problematic for bosonic open string field theory such a construction 
could in principle be completed successfully in open superstring field 
theory. [5]

Another possible clue to the cosmological constant problem resides in the 
effective field theory view of gravity. Another way of looking at the 
cosmological constant problem is that our effective field theory low energy 
description of gravity at standard model energy scales must be missing 
something. If string theory is the correct description of nature this means 
some unexpected stringy effect must survive at energies far below the string 
scale, i.e., at a scale where general relativity is normally considered a 
completely accurate effective field theory. Therefore, the field theory 
limit of string theory may warrant closer inspection.

In this paper the above two considerations are combined to see if they can 
shed light on the cosmological constant problem. A completely heuristic 
approach is taken. It is simply assumed that a detailed construction of a 
composite graviton can eventually be done in the context of open superstring 
field theory, and the closed string sector arises from the open string 
loops. The approach is then backward in the sense that it starts from where 
we are (standard model energy scales), rather than starting from the string 
scale and working down as is normally done, but without adopting a field 
theory limit of the underlying string theory. The only other input is the 
open string spectrum we observe (i.e., the standard model particles). 
Although motivated by string field theory the stringy aspects incorporated 
are essentially conventional world sheet ideas. In particular the conformal 
invariance of the world sheet is employed to extract the graviton pole from 
the nonplanar equivalent of the open string loops along with a UV to IR map 
between the observed open string sector and a closed string composite 
sector/graviton sector. As a result of this map a composite graviton UV 
energy cutoff is related to the IR cutoff of the open sector vacuum loops. 
This IR cut off is the estimated Compton wavelength of the lightest standard 
model particle having the requisite loop structure, the neutrino. Reasonable 
estimates of the neutrino mass are in the order of 10$^{-3}$ electron volts, 
providing a graviton UV cut off of the same order of magnitude as the 
estimated cosmological constant.\footnote{ A connection between the neutrino 
mass and cosmological constant has also recently been proposed in a 
composite graviton model in [12], however, the specific model is quite 
different.} A phenomenological modification to the graviton propagator is 
presented which incorporates this cut off. 

Although this simple heuristic model does not solve the cosmological 
constant problem per se since it needs the neutrino mass as an input, it has 
some suggestive qualitative features which should survive in a more detailed 
string field theory construction. In particular, it suggests the 
cosmological constant problem may simply be another aspect of the hierarchy 
problem, or at least the aspect of the hierarchy problem related to the 
light fermion masses, which is hopefully more tractable. Other potentially attractive features of 
this model are also discussed. 

2. THE STRINGY COMPOSITE GRAVITON

As noted above a composite graviton constructed from the loops of an open 
string sector will simply be assumed.\footnote{ The term composite graviton 
is really a misnomer in this context and falsely suggests that the no go 
theorem of [13] should apply. However, the no go theorem of [13] assumes a 
completely consistent underlying field theory without a spin two state in 
which a spin two particle then emerges as a bound state. In contrast, open 
string theory without a closed sector is inconsistent and a ``composite'' 
closed sector is added at a fundamental level to make a consistent theory. 
The no go theorem is therefore not applicable. Since this is a purely 
stringy effect a ``stringy composite graviton'' as discussed here should be 
clearly distinguished from other ``composite graviton'' models which have 
appeared for many years.} This is believed to be a not unreasonable 
assumption since the current status of open string field theory is 
suggestive that a complete construction of this type is possible. [5] There 
are other motivations as well and a general discussion of possible 
motivations for a composite closed string are found in [6]. [5] and [6] 
contain many additional references. 

Ideally we would like to make detailed contact with the existing open string 
field theory models for further guidance. However, the current state of 
development is not really able to give much insight beyond suggesting a 
consistent open string theory with a composite closed sector is possible. In 
particular, superstring field theory does not yet exist in a form which can 
describe exactly how the closed string poles arise or what the relevant 
boundaries of moduli space are. Also, a composite closed sector may arise 
from another nonperturbative formulation of string theory, e.g., a matrix 
model, that will nonetheless presumably keep the key features of world sheet 
theory. Therefore, we instead fall back on world sheet ideas as a guide. 
Since we are only looking for a qualitative guide, it turns out that the 
only key stringy features we will need beyond assuming a composite model are 
world sheet conformal invariance and the world sheet modular 
parameter.\footnote{ The basic world sheet ideas used are not new and 
essentially just follow the discussion in section 4 of [6].}

We are assuming open sector loops in some sense define the closed sector. 
Our interest is thus in the (super)string annulus diagram describing open 
loop amplitudes and a nonplanar cylinder diagram for the closed 
sector.\footnote{ We know from standard arguments involving unitarity that 
some type of closed diagram with a pole structure is needed to make our open 
string theory consistent. See, e.g., section 3.1 of [7], vol. 1.} We will 
simply assume the amplitude structure is the same. The amplitude structure 
is determined by the integral over the world sheet modular parameter, which 
we will take as the cylinder modular parameter t, the ratio of cylinder 
circumference to length. The t $\to \quad \infty $ limit is the IR part of the 
open sector annulus. This limit is not problematic, however, since such IR 
limits are well understood. At the other limit, t $\to $ 0, the open string 
annulus amplitude blows up and is potentially dangerous. However, this limit 
is also conventionally interpreted as a soft IR limit, but of the closed 
strings, corresponding to arbitrarily long distance propagation of long 
skinny tubes. That is this edge of the world sheet integration has a 
singularity which can be represented (if we were to take a field theory 
limit) in the form of a field theory massless particle pole:
\[
D(k)\sim 1/k^2
\]
which allows the dangerous UV sector of the open loop to be reinterpreted as 
an infrared effect. Therefore, the short distance open string UV vacuum 
fluctuations naturally correspond to long distance propagation, the IR 
region of the closed string. The large open loops in turn correspond to 
short distance closed loop propagation, the UV region of the closed string. 
Therefore, we have the limits of the same amplitude structure mapped to two 
different regions of the open and closed sectors; the t $\to \quad \infty $ 
limit corresponds to open sector IR, closed sector UV, and the, t $\to $ 0 
limit corresponds to open sector UV, closed sector IR. 

At this point, a brief discussion of some recent technical results of open 
string field theory is appropriate since they appear to resolve a potential 
conflict with this simple qualitative picture of the UV to IR 
correspondence. In open string field theory the pole of the closed sector 
actually arises in a different way. It still occurs at the t $\to $ 0 limit, 
but with the closed tubes pinching off at a point and effectively both t and 
the string length L going to zero at the pole. [15] However, this type of 
pinching off actually seems to be an inconsistent description of the closed 
sector pole and is actually signaling an anomaly and loss of conformal 
invariance. [5] Proper incorporation of the closed sector appears to be able 
to cancel the anomaly, at least at one loop. [5] In that sense the closed 
sector is like a Wess-Zumino term; certainly a different way of looking at 
the composite graviton. On the other hand if this anomaly cancellation program can be successfully completed, with conformal invariance intact 
the world sheet view should again have validity, at least in a qualitative 
sense. That is, the spacetime string field theory view with an anomaly and a 
compensating closed sector, and the world sheet view of the closed sector as 
a well behaved conformal limit of the modular parameter t, may be equally 
valid ways of looking at the same composite closed sector. Certainly, with 
all the insights that have come from the world sheet view it does not seem 
unreasonable to retain the general world sheet view and retain the 
qualitative features of the UV to IR correspondence discussed above.

Now we turn to our second input which is the observed standard model 
particles and energies which we want to impose while retaining our stringy 
view of the open and closed sectors with a composite closed sector. Of 
course there is no reason to assume our simple open-closed correspondence at 
a highly symmetrical string scale starting point should survive the ``field theory limit'' (as well as a complicated combination of compactification and symmetry breaking effects). However, the goal here is to simply assume some of this stringy open- closed 
amplitude correspondence does survive at low energy and see if the open 
sector can provide some clues to the closed sector. 

As a first guess, it could be reasonably expected that the edges of moduli 
space dominate the low energy limit for both the planar (annulus) and nonplanar (cylinder) 
amplitudes. Therefore, we should look there for signals that stringy effects can 
survive at low energies in the closed sector, using the open sector as a 
guide. Therefore, let us initially ignore the interior of moduli space and 
just consider the edges.

First of all, the correspondence at the t $\to $ 0 part of moduli space appears qualitatively 
unchanged at low energies in the open and closed sectors. The UV sector of 
the standard model still blows up which we may still equate to the pole in 
the closed sector as before. Therefore, a massless composite graviton correlating to 
the open sector loops could still be a consistent part of the theory we see at low energy. The other edge of moduli 
space is defined by the t $\to \quad \infty $ limit. This limit of the annulus 
is defined by the largest color/charge singlet loops; the IR edge of the 
standard model (open string) sector. This edge clearly has changed. This 
edge is defined by the mass of the lightest standard model particle with a 
color/charge singlet vacuum loop diagram, namely the neutrino. We will 
assume the amplitude structure correspondence has remained the same at this 
edge. A corresponding change in the closed sector at this edge thus requires 
a cut off of the cylinder at a finite value. In the closed sector this does 
not affect the massless pole structure, which comes from the other edge of 
moduli space, but it does impose a UV cut off. Therefore, since our cylinder 
diagram is the graviton propagator we get a UV graviton energy cutoff which 
corresponds to the neutrino mass edge of moduli space.
Therefore, even 
if we assume that only the correspondence of the open and closed amplitudes 
at the edges of moduli space survives at low energy we get a departure from 
a conventional low energy effective field theory of gravity and get a UV 
gravity cutoff. This is potentially the low energy stringy effect we were 
looking for.

However, at this point a potential problem presents itself. Why should the 
neutrino, or any standard model fermion, have anything to do with gravity or 
the cosmological constant problem? The graviton is a spin two mode of the 
superstring and the fermion vacuum loops are spin zero. Why should the 
latter have anything to do with the former? Actually this is not an 
inconsistency. First of all, the cosmological constant problem in a Feynman 
diagram sense is related to the graviton tadpole. To avoid violation of 
Lorentz invariance this tadpole must be the scalar mode of the graviton. 
Viewed from a classical field theory sense (i.e., linearized general 
relativity), while the scalar mode of the graviton can be globally gauged 
away in the absence of matter, it can only be gauged away locally in the 
presence of matter. Since the vacuum energy/cosmological constant is 
effectively the same as a matter term, the scalar mode cannot be gauged away 
globally, only locally, if a cosmological term is present. Therefore, the 
cosmological constant problem inherently is rooted in the scalar component 
of the graviton.$^{ }$\footnote{ For a brief early discussion of the 
graviton tadpole and cosmological constant see [9] and more recently [10].} 
Also, the scalar component of the graviton includes the dilaton which is 
also present in the closed sector. As in conventional string theory 
approaches the dilaton couples to the effective gravitational constant (G). 
Therefore, the neutrino edge of moduli space can indeed be relevant to both 
the cosmological constant and the full effects of gravity.

This leads to another issue which is whether we can find a possible open 
sector equivalent of the spin two component of the graviton. This spin two 
mode presumably could only correlate to the nonAbelian gauge bosons of the 
open sector since fermion vacuum loops cannot create a spin two 
mode.\footnote{ The apparently contrary claims of [12] have not been 
carefully examined, however.} Actually such a nonAbelian gauge boson 
$\leftrightarrow $ gravity correspondence has been identified and 
extensively studied (in a different context) and indeed graviton vertices 
can be consistently mapped to direct products of nonAbelian gauge 
amplitudes. [11] 

Therefore, in principle both the scalar/dilaton and spin two modes of the 
closed string sector can be accommodated in the open string sector. One way 
to look at the resulting graviton structure is that the compositeness we 
have added in, along with the stringy UV cutoff, is effectively encoded in 
the dilaton, while the spin two mode incorporates the standard massless pole 
structure. This seems reasonable since the dilaton reflects the freedom to 
rescale the closed loop size and it could logically encode all the effective 
graviton size of the composite graviton.

Another way of looking at the reason a stringy effect remains at low 
energies is the different nature of the field theory limit in a composite 
model. In the field theory limit the string length goes to zero. In a 
conventional closed string model the closed string loop circumference goes 
to zero and the closed string shrinks to a point. In the composite model the 
length of the open string still goes to zero but the closed loop remains 
finite. Only by turning off quantum effects could the loop size be affected, 
but then it disappears rather than shrink to a point. This highlights the quantum origins of the closed sector in such a composite model. This has already been evident for some time from the appearance of $\hbar $ in the closed sector coupling constant in the open-closed string field theory of [8] and the composite string field theory model of [4].   

The application of our stringy UV gravity cutoff to the cosmological 
constant problem is straightforward. To make a na\"{\i}ve estimate of the 
value of the cosmological constant we can simply look at the highest energy 
gravitons as these will give the biggest contribution. As discussed above 
this correspond to the UV cutoff or the largest loops of the open sector, 
which corresponds to the neutrino mass edge of moduli space. The neutrino 
mass has been estimated to be on the order of 10$^{-3}$ eV. This is of the 
same order of magnitude as the current estimates of the dark energy or 
effective cosmological constant. Although this could be a coincidence it is 
at a minimum very suggestive.

Assuming from the above that there are stringy effects missing from the 
classical gravity sector, and in particular in the gravity UV region, we can 
try to make a phenomenological adjustment to the graviton propagator. We can 
simply modify the form of the propagator to include the UV edge of moduli 
space as follows:
\[
D(k)\sim F(k)_{UV} (1/k^2)
\]
where the graviton tensor indices are suppressed and F(k)$_{UV}$ has a 
momentum dependence with a cutoff at the UV energy edge of the 
moduli space as signaled by the neutrino. F(k) cannot, however, affect the 
simple pole structure of the propagator to avoid violating unitarity. For 
example, the following F(k) may provide the desired cutoff without affecting 
the basic long range graviton pole: 
\[
F(k)_{UV} =1/(A+Be^{k/C-1})
\]
where C $\sim $ 10$^{-3}$ eV and A and B normalize F(k) to 1 outside of our 
cutoff regime (the possibility that B could include a very slow k dependence 
is discussed in the next section). More complicated forms of F(k) are of 
course possible and cannot be ruled out other than by experiment or a 
detailed model.

At a Lagrangian level the modifications due to the composite graviton are 
straightforward in principle. The conventional string theory derived 
effective Lagrangian density for gravity is of the following form:
\[
L=\frac{1}{2k^2}\sqrt {\left| g \right|} e^{-2\phi }(R+4\partial _\mu \phi 
\partial ^\mu \phi +\cdot \cdot \cdot )
\]
where $\phi $ is the dilaton, R is the Ricci scalar and {\ldots} indicates 
higher order terms. The gravitational constant G is determined via 
e$^{-2\phi }$/k$^{2} \quad \sim $ G. As discussed above, the light fermion/UV 
gravity cutoff corresponds to the scalar sector of the graviton which mixes 
with the dilaton. Therefore, this cutoff can be simply incorporated into the 
standard Lagrangian by incorporating this into the dilaton potential. One 
difference, however, is the relation between the dilaton in the open and 
closed sectors. In the present model the dilaton is a composite so the 
manner in which the dilaton mixes with the open sector should in general 
differ from conventional derivations of the dilaton-matter coupling. This 
composite dilaton could potentially lead to a dilaton which decouples from 
matter or more generally a dilaton potential similar to that proposed in the 
decoupling mechanism of [14] which has a number of desirable features. This 
could also remove the direct link between time variation of standard model 
coupling constants and the dilaton potential relevant to the gravitational 
sector which could actually lessen some experimental constraints within the 
present model.

3. FURTHER POSSIBLE STRINGY CORRECTIONS

The above modification of the closed string/gravity sector only incorporates 
the modified neutrino defined UV edge of the moduli space. A question 
obviously arises whether or not the rest of the moduli space other than the 
neutrino edge further modifies the graviton propagator. The open string 
sector/standard model has a complicated mass structure for particles that 
have singlet loop graphs which could potentially be parts of the same ``edge'' of moduli space and signal effects in the closed 
sector. It is at least conceivable that this additional structure of the 
standard model could be encoded in the graviton/dilaton without violating 
observational tests of GR. For example, the details of the UV to IR 
correspondence of this ``edge'' of moduli space (which will also be affected by 
compactification) could put these other variations in gravity so far in the 
IR long distance regime that they are not in conflict with experimental 
tests of GR; possibly even outside the observable cosmological horizon. 

Therefore, whether or not observable, a modification to the graviton 
propagator of the following generic form is possible:
\[
D(k)\sim \sum\limits_i {F(k)_i } (1/k^2)
\]
where the sum is over the mass sectors of the standard model. As one 
example, we could have:
\[
F(k)_i =1/(A_i +B_i e^{k_i /C_i -1})
\]
where the C$_{i}$ are the respective mass cutoffs and A$_{i}$ and B$_{i }$ 
are respective normalizations to adjust the relative contribution of the different standard model mass sectors. In terms of the gravitational coupling this 
would translate via the dilaton to an unusual step type variation in G. Since the details of the open-closed correspondence are unknown a more generic form of F(k) could also include a slow 
``conformal mapping factor'' by giving B a slow
depedendence on k, but which cannot introduce any cuts into the propagator. 

If these modifications are within observable distance these contributions to 
the graviton propagator and G must be small relative to the neutrino scale 
UV cutoff effect to not conflict with experiment. However, this is not 
unreasonable to expect from the results of compactification, the relative 
effect of the interior of the moduli space to the moduli edge contributions, 
and the details of the UV to IR map. It is therefore interesting to consider 
such small but nonzero long distance modifications for galactic or 
cosmological constraints. In this regard, the data supporting MOND as an 
alternative to dark matter suggest that some long range modifications of 
this type are not ruled out experimentally. The k dependence of the graviton 
propagator could naturally tie in to an acceleration threshold dependence as 
required by MOND. The unusual step type dependence could be an 
advantage in this setting. In any event if some component of this part of 
the moduli space is nonzero in the long range graviton propagator, no matter 
how small, it would provide an interesting new way to look for clues to this open-closed correspondence in the large scale gravitational dynamics of the universe.

4. THE STRINGY GRAVITON AND INFLATION

One generic problem with any composite graviton as a solution to the 
cosmological constant problem is that the effective cosmological constant 
not only needs to be small but nonzero now but must also have been large 
during the early universe. If the graviton has a high energy cut off due to 
a composite structure it should be irrelevant in the early universe but 
inflation requires a large gravitational negative energy density. Nonetheless, the 
present model is not incompatible with strong gravitational effects at early 
times. This follows because the graviton UV cutoff is tied to the light but 
nonzero lightest fermion mass scale. This fermion mass scale doesn't emerge until 
after a phase transition well along in the evolution of the universe 
(relatively speaking). No limitation on the t $\to \infty $ edge of the 
modular integration is present before then. Therefore, during the early 
inflation era the above discussed UV cutoff is not present and $\Lambda $ is 
not limited by this cutoff.

Indeed the early universe $\Lambda $ could have its normally predicted high 
value expected from vacuum quantum effects which alone could potentially 
drive inflation and potentially replace the inflaton. A mass generating phase transition could trigger the end of inflation. It would certainly be 
desirable to avoid the at present essentially ad hoc inflaton field. On the other hand such a model has a number of potential hurdles. The viability of this
type of scenario would depend on the details of the initial string theory 
vacuum state and how supersymmetry could be incorporated into such a model while allowing a large $\Lambda $ at early times. This would presumably require either a natural string theory ground state with unbroken supersymmetry and large $\Lambda $ or a mechanism to break supersymmetry in the early universe and start inflation. Also, how this approach could give all the necessary features of inflation without an inflaton potential is not clear. Nonetheless, there conceivably is room for an 
inflation scenario with a composite graviton and without an inflaton. At a minimum the 
ability to accommodate both a large early effective $\Lambda $ and a small 
nonzero $\Lambda $ now is an important feature for compatibility with 
inflation. It seems likely, however, that the correct description of the composite graviton's role in the early universe will require a completed open string field theory or other nonperturbative formulation and our simple world sheet view will be inadequate to answer these questions.  

5. OTHER POSSIBLE FEATURES

At the simplest level the composite graviton model discussed above simply 
imposes a UV cutoff to gravity. Even limited to this apparently simple 
phenomenological modification, however, the model potentially has desirable 
features other than shedding light on the cosmological constant problem. A 
UV cutoff could remove the singularity problems which plague classical 
general relativity. Since the composite graviton is inherently stringy in 
nature it preserves the string theory insights into black hole entropy and 
could provide further insights in this direction. Also, the UV to IR map is 
suggestive of AdS/CFT and this in turn suggests that a holographic 
interpretation could be found. Also, the connection between the hierarchy 
problem and the cosmological constant problem which emerges naturally is 
quite appealing. Therefore, even at the simplest level, a stringy composite 
graviton has far reaching implications.

Also, the model has testable features. Although the UV cutoff is consistent with present experimental limits, variations in G below the neutrino mass scale should be detectable in the not too distant future. This basic signature is shared with the ``fat graviton'' model of R. Sundrum [3]. In addition, due to the relation of the effective $ \Lambda $ to mass generation of the lightest fermion (neutrino) other even more unique signatures are possible.\footnote{ The specifics of such potentially testable signatures are currently being explored further.}

6. DISCUSSION

The foregoing is a highly qualitative argument for a possible way out of the 
cosmological constant problem in the context of the closed string sector 
emerging out of a purely open string theory. At its most basic the argument 
is that we should not simply collapse the graviton to a simple field theory 
pole in an effective field theory approach to gravity as this ignores 
stringy aspects of the graviton. Also, the appearance of the lightest 
fermion (neutrino) mass apparently at the same order of magnitude as the 
observed dark energy is certainly at least suggestive that the observed 
standard model can give insight to this missing stringy structure of the 
graviton. Therefore, it seems a reasonable argument can be made for a 
graviton with a stringy composite structure.

REFERENCES:

[1] For a review see S. Weinberg, Rev.Mod.Phys.61:1-23(1989).

[2] See L. Susskind, hep-th/0302219.

[3] E.g., A. Zee, hep-th/0309032; R. Sundrum, hep-th/0306106, and 
hep-th/031025; N. Arkani-Hamed, S. Dimopoulos, G. Dvali, G. Gabadadze, 
hep-th/0209227.

[4] W. Siegel, Phys.Rev.D49:4144-4153(1994). 

[5] I. Ellwood, J. Shelton, W. Taylor, JHEP 0307:059,2003.

[6] J. Polchinski, What is String Theory?, Les Houches 1994 Summer School 
Lectures, eds. F. David, P. Ginsparg {\&} J. Zinn-Justin, pp. 287-422 
(North-Holland 1996).

[7] J. Polchinski, STRING THEORY. VOL. 1 and 2 (Cambridge University Press 
1998).

[8] B. Zwiebach, Phys.Lett.B256:22-29(1991).

[9] M. Veltman, in Les Houches 1976, ed. R. Balian and J. Zinn-Justin, 
(North --Holland 1976), pages 265-326.

[10] A. Adams, J. McGreevy, E. Silverstein, hep-th/0209226.

[11] Z. Bern, gr-qc/0206071.

[12] J. W. Moffat, gr-qc/0309125.

[13] S. Weinberg, E. Witten, Phys.Lett.B96:59(1980)

[14] T. Damour, A. M. Polyakov, Gen.Rel.Grav.26:1171(1994).

[15] C. B. Thorn, Phys. Rep. 175: 1(1989).

\end{document}